\documentclass[twocolumn,showpacs,preprintnumbers]{revtex4} 
\def\to{\rightarrow}
\def\gev{\mbox{GeV}}

\def\mev{\mbox{MeV}}
\def\tev{\mbox{TeV}}

\def\CQG{{\it Class. Quantum Gravity} }

\def\NP{{\it Nucl. Phys.} }
\def\PL{{\it Phys. Lett.} }
\def\PR{{\it Phys. Rev.} }
\def\PRL{{\it Phys. Rev. Lett.} }

\def\frac#1#2{{\textstyle{{#1}\over {#2}}}}

\def\lsim{\mathrel{\rlap{\lower4pt\hbox{\hskip1pt$\sim$}}
    \raise1pt\hbox{$<$}}}
\def\gsim{\mathrel{\rlap{\lower4pt\hbox{\hskip1pt$\sim$}}
    \raise1pt\hbox{$>$}}}
\def\sqr#1#2{{\vcenter{\vbox{\hrule height.#2pt
         \hbox{\vrule width.#2pt height#1pt \kern#1pt
         \vrule width.#2pt}
         \hrule height.#2pt}}}}

\def\beq{\begin{equation}}
\def\eeq{\end{equation}}
\def\beqa{\begin{eqnarray}} 
\def\eeqa{\end{eqnarray}}

\begin{document}
\preprint{DF/IST-11.2001, FISIST/13-2001/CFIF }

\title{ $N=1$ Supergravity Chaotic Inflation in the Braneworld Scenario}

\author{M. C. Bento}
\altaffiliation[Also at] { Centro de F\'{\i}sica das 
Interac\c c\~oes Fundamentais, Instituto Superior  T\'ecnico,
Av. Rovisco Pais 1, 1049-001 Lisboa, Portugal}
\email{bento@sirius.ist.utl.pt}
\author{O. Bertolami}
\email{ orfeu@cosmos.ist.utl.pt}
\affiliation{Departamento de F\'\i sica,
 Instituto Superior  T\'ecnico\\
Av. Rovisco Pais 1, 1049-001 Lisboa, Portugal}

\date{\today}

\begin{abstract}
We study a N=1 Supergravity  chaotic inflationary model,
in the context of the braneworld scenario. It is shown that 
successful inflation and reheating consistent with phenomenological 
constraints can be achieved via  the new terms in the Friedmann equation 
arising from  brane physics. Interestingly, the model satisfies 
observational bounds  with sub-Planckian field values, 
implying that chaotic inflation on the brane is free from the well 
known difficulties associated with the presence of higher order 
non-renormalizable terms in the superpotential. A bound on  the mass scale
 of the 
fifth dimension, $M_5 \gsim 1.3 \times 10^{-6}~M_P$, is obtained from the 
requirement that the reheating 
temperature be  higher than the temperature of the 
electroweak phase transition.
  
\vskip 0.5cm
 
\end{abstract}

\pacs{98.80.Cq }

\maketitle
\section{Introduction}

Chaotic inflationary models \cite{Linde}
stand out for their simplicity and  fairly natural initial
conditions for the onset of inflation. These features are
particularly appealing in the context of supergravity and superstring
theories, where the natural scale for fields is the Planck scale. In the
context of $N=1$ Supergravity, however, realizations of chaotic inflation in
minimal and in $SU(1,1)$ theories are somewhat special as to ensure
sufficient inflation \cite{Goncharov,Goncharov1}.
Furthermore, chaotic inflation requires super-Planckian field values both
to ensure sufficient inflation and the correct amount of 
Cosmic Microwave Background
(CMB) anisotropies, in which case  higher order non-renormalizable 
terms in the superpotential would completely dominate the dynamics  since
no well motivated symmetry is known to prevent them. Assuming that this
problem can be circumvented somehow,  one finds that, in order to accommodate
in a satisfactory way the bounds on the reheating temperature and energy
density fluctuations, a two-scale chaotic inflationary sector
is required \cite{Bento1}, in contrast with the situation found 
in $N=1$ Supergravity 
new inflationary type models, where a single scale suffices \cite{Ross,Holman}.

In this work, we show that specific features of the braneworld scenario allow
for the abovementioned difficulties with higher order non-renormalizable 
terms to  be quite naturally avoided and that current observational contraints
can be accounted for with a single scale at the superpotential level.

Higher dimensional superstring motivated cosmological 
solutions suggest that matter
fields that are related to open string modes lie on a lower dimensional brane
while gravity propagates in the bulk \cite{Polchinski}. It is striking that,
in these scenarios, extra dimensions are not restricted to be small
\cite{Antoniadis} and that the fundamental $D$-dimensional scale, 
$M_D$, where $D=4+d$, can be considerably smaller than the 4-dimensional
Planck scale. In
this work, we shall consider the $D=5$ case. If one assumes that Einstein
equations with a negative cosmological constant hold (an anti-De-Sitter 
space is required) in $D$-dimensions and that matter
fields are confined to the $3$-brane then the $4$-dimensional Einstein equation
is given by \cite{Shiromizu}:

\beq 
G_{\mu \nu} = - \Lambda  g_{\mu \nu} + {8 \pi \over M_P^{2}} T_{\mu \nu} 
+ \left({8 \pi \over M_{5}^3}\right)^2 S_{\mu \nu} - E_{\mu \nu}~,
\label{eq:gmunu}
\eeq
where $T_{\mu \nu}$ is the energy-momentum on the brane, $S_{\mu \nu}$ is a 
tensor that contains  contributions that are quadratic in
$T_{\mu \nu}$ and $E_{\mu \nu}$ corresponds to the projection of the 
5-dimensional Weyl tensor on the 3-brane  (physically, for a perfect fluid, 
it is associated to non-local contributions to the pressure and energy 
flux). 
The $4$-dimensional cosmological constant is related to the $5$-dimensional
cosmological constant and the $3$-brane tension, $\lambda$, as

\beq
\Lambda = {4 \pi \over M_5^3} \left(\Lambda_5 + {4 \pi \over 3 M_5^3}~ 
\lambda^2 \right)~,
\label{eq:lam}
\eeq
while the Planck scale is given by

\beq
M_P = \sqrt{{3 \over 4 \pi}} {M_5^3 \over \sqrt{\lambda}}~~.
\label{eq:MP}
\eeq

In a cosmological setting, where the $3$-brane resembles our Universe and 
the metric projected onto the brane is an homogeneous and isotropic flat
Robertson-Walker metric, the generalized Friedmann equation has the
following form \cite{Binetruy}:

\beq
H^2 = {\Lambda \over 3} +  \left({8 \pi \over 3 M_P^2}\right) \rho 
+ \left({4 \pi \over 3 M_5^3}\right)^2 \rho^2 + {\epsilon \over a^4},
\label{eq:H2}
\eeq
where $\epsilon$ is an integration constant arising from $E_{\mu \nu}$.
During inflation, the last term in (\ref{eq:H2}), the ``dark radiation'' 
term, rapidly vanishes and will be disregarded
hereafter. Moreover, observations require that the cosmological constant is
negligible in the early Universe, meaning that $\Lambda_5$ and the brane
tension were fine tuned, $\Lambda_5 \simeq - 4 \pi \lambda^2/3 M_5^3$. 
The Friedmann equation  can then be written in the following way:

\beq
H^2 = {8 \pi \over 3 M_P^2} \rho \left[1 + 
{\rho \over 2 \lambda}\right]~~.
\label{eq:H22}
\eeq
Notice that the new term in $\rho^2$ is dominant at high energies, compared
to $\lambda^{1/4}$, but quickly decays at lower energies. Requiring the new 
term to be  sub-dominant during nucleosynthesis, implies  that $\lambda \gsim
(1~\mev)^4$ and, therefore \cite{Cline}:

\beq
M_5 \gsim \left({1~\mev \over M_P}\right)^{2/3} M_P = 10~\tev~~.
\label{eq:M5}
\eeq 

A much more stringent bound can be obtained in models where the fifth
dimension is infinite as it yields an extra contribution $M_5^6\lambda^{-2}
r^{-2}$ to 
Newton's gravitational force, which should be, of course, small beyond scales 
 $r \gsim 1~mm$ (see eg. Ref. \cite{Floratos} and references therein). 
From Eq. (\ref{eq:MP}), it follows that $M_5 > 10^5~\tev$. We shall see 
in the ensuing discussion that 
cosmological considerations intrinsic to supersymetric cosmology can set even 
more stringent bounds on $M_5$.

In what follows, we shall discuss the effect of the new term in the Friedmann 
equation on the slow roll-over parameters and show that 
it allows for a single scale chaotic
inflationary model, in the context of $N=1$ Supergravity. Moreover, we study 
how parameters in the superpotential and $\lambda$ 
are affected by constraints on the 
magnitude of the energy density perturbations required to explain the
anisotropies in the CMB  radiation observed by COBE as well as on the 
reheating temperature.
In fact, in  supergravity cosmological models one has to ensure that the 
reheating temperature does  not
exceed $T_{RH}~\lsim~ 2.5\times  10^8 (100~\gev/m_{3/2})~\gev$  \cite{Ellis}, 
in order not to generate an over-abundance of gravitinos and the ensuing
photo-dissociation of light elements at  nucleosynthesis. In the 
context of superstring cosmology, inflationary models have to face further
problems such as the fate of the dilaton and moduli fields, the so-called
post-modern Polonyi problem \cite{Coughlan}.

We also  show that the observational quantities arising from 
the model originate in sub-Planckian field regimes and, hence, 
higher dimensional non-renormalizable
operators in the superpotential are not relevant. 

\section{The Model}

We shall assume that the inflaton is the scalar component of a gauge
singlet superfield, $\Phi$, in the hidden sector of the theory. We start 
by  splitting the superpotential in a supersymmetry-breaking, a
gauge and an inflationary sector:

\beq
\label{eq:W}
W= P + G + I~.
\eeq

The $N=1$ supergravity theory describing the interaction of gauge singlet 
fields is specified by the K\"ahler function, in terms of which the scalar 
potential is given by

\beq
\label{eq:V}
V={1\over 4} e^K \left[ G_a (K^{-1})_b^a G^b -3 | W |^2\right]
\eeq
where $G_a=K_a W + W_a$, and the indices $a,\ b$ denote derivatives 
with respect to the chiral superfields, $\Phi$.
We consider the minimal choice for  the K\"ahler function in $N=1$ 
Supergravity:

\beq
\label{eq:K}
K(\Phi, \Phi^\dagger)= \Phi^\dagger \Phi~, 
\eeq
corresponding to canonical kinetic energy for the scalar fields.
With this choice, the scalar potential for the inflaton field, $\phi$, is
obtained from the superpotential $I(\Phi)$ as 

\beq
\label{eq:V2}
V(\phi)=e^{{| \phi |}^2\over M^2} \left[ \left|{{\partial I}\over
{\partial \Phi}} + 
{\Phi^\star I\over M^2}\right|^2- 3 {{| I |}^2 \over M^2}\right]_{\Phi=\phi}~~,
\eeq
where  $M=M_{P}/\sqrt{8 \pi}$. 
 
Requiring the cosmological constant to vanish and that supersymmetry 
remains unbroken at the minimum of the
potential,  $\Phi=\Phi_o$,  leads to
the following constraints on the superpotential:

\beq
\label{eq:phi0}
I(\Phi_o)={{\partial I}\over{\partial \Phi}}(\Phi_o)=0~~.
\eeq 

Consider the simplest form for the  superpotential $I$ 
which satisfies the above conditions

\beq
\label{eq:I}
I(\Phi)= \mu (\Phi - \Phi_0)^2~~,
\eeq
where  $\mu $ is a  mass  parameter that determines 
the energy scale for inflation. Hereafter, we set $\Phi_0=0$. 
The relevant part of the inflaton
potential (along the real $\phi$ direction) is then given by:

\beq
\label{eq:pot}
V(\phi) = \mu^2  \left(4  \phi^2 +  {\phi^4\over M^2} + {\phi^6 \over M^4}
 \right) e^{\phi^2\over M^2}~~.
\eeq

Consistency between the  slow-roll approximation  and  the
full evolution equations requires that there are  constraints on the slope and
curvature of the potential. One can define two slow-roll parameters 
\cite{Maartens}

\begin{eqnarray}
\label{eq:epsilon}
\epsilon &\equiv&{M_{P}^2 \over 16\pi} \left( {V' \over V}
\right)^2  {1+{V/ \lambda}\over(2+{V/\lambda})^2}
~~,\\
\label{eq:eta}
\eta &\equiv& {M_{P}^2 \over 8\pi} \left(
{V'' \over V} \right)  {1 \over 1+{V/ 2 \lambda}}~~.
\end{eqnarray}
Notice that   both parameters are 
suppressed by an extra factor $\lambda/V$ at high energies and that, at low
energies, $V\ll\lambda$, they reduce to the standard
form. The end of inflation will take place for a field value, $\phi_f$,
such that

\beq
\label{eq:phif}
{\rm max}\{\epsilon(\phi_f),|\eta(\phi_f)|\}= 1~~.
\eeq

The number of e-folds during inflation is given by $N =
\int_{t_{\rm i}}^{t_{\rm f}} Hdt$, which  becomes \cite{Maartens}
\beq
\label{eq:N}
N \simeq - {8\pi  \over M_{P}^2}\int_{\phi_{\rm i}}^{\phi_{\rm
f}}{V\over V'} \left[ 1+{V \over 2\lambda}\right]  d\phi~~,
\eeq
in the slow-roll approximation. We see that, as a result of  the modified
Friedmann equation at high energies,
the expansion rate is increased  by a factor $V/2\lambda$,
allowing for a smaller initial inflaton field value, $\phi_{\rm i}$, for a  
given number of e-folds, which is crucial for achieving the 
goal of obtaining sufficient inflation with sub-Planckian field values 
in our model.

\section{Constraints from Reheating}

After inflation,  the field $\phi$ releases its energy via the coupling 
to fields in the other sectors in (\ref{eq:W}), thus reheating the Universe. 
Since the inflaton is hidden from the other sectors of the theory, it
couples to lighter fields with gravitational strength of the order ${\mu \over M}$. 
At minimum, the inflaton field has a mass  

\beq
\label{eq:m}
m_\phi= 2\sqrt{2} ~\mu~~,
\eeq 
leading to a decay width

\beq
\label{eq:Gam}
\Gamma_\phi \simeq {{m_\phi}\over{(2\pi)^3}}~ \left(\mu \over M\right)^3~~,
\eeq
and a reheating temperature

\beqa
\label{eq:TRH}
T_{RH}&\simeq &\left({30 \over \pi^2 g_{RH}}\right)^{1/4} 
\sqrt{M \Gamma}\nonumber \\
&\simeq &{2 \over \pi^2}\left(\sqrt{15 \over g_{RH}}{\mu^3 \over M}\right)^
{1/2}~~,
\label{eq:TRH2}
\eeqa
where $g_{RH}$ is the number of degrees of freedom at $T_{RH}$.

As already mentioned, a quite severe upper bound on $T_{RH}$ comes from the
requirement that gravitinos are not abundantly regenerated in the
post-inflationary reheating epoch. Indeed, once regenerated beyond a
certain density, stable thermal
gravitinos would dominate the energy density of the Universe or, if
they decay, have disruptive effects on nucleosynthesis, causing light element
photo-dissociation and distortions in the CMB.
Avoiding these difficulties implies in the following bounds \cite{Ellis}:

\beq
\label{eq:bca}
T_{RH}~\lsim ~2\times 10^9,~6\times 10^9~\gev
\eeq
for $m_{3/2} = 1,~10~\tev$. 
For our model, demanding that $T_{RH}$ be less than $2\times 10^9~\gev$ 
leads, for $g_{RH} = 150$, to a limit on parameter $\mu$ 

\beq
\label{eq:bd}
\mu~\lsim~3.7\times 10^{-6}~M~~,
\eeq
which coincides with the bound  obtained in Ref. \cite{Bento1} for $SU(1,1)$ 
Supergravity since, in either case,  reheating is naturally 
controlled by the lowest order quadratic term in the superpotential.

We have checked that the above bound also ensures that gravitino production 
via inflaton decay is sufficiently suppressed, as in \cite{Bento1,Ross}.

Naturally, one could contemplate other more specifical 
reheating mechanisms. Parametric 
resonance \cite{Kofman} seems to be unimportant given that 
$m_{\phi} >> \Gamma_\phi$ and the absence of bilinear type couplings in 
supergravity. On the other hand, anharmonic terms coupling  the inflaton 
with other chiral superfields can be an alternative  route for reheating 
\cite{Bertolami}. In the braneworld scenario, gravitational particle 
production has been discussed for an exponentially decaying inflaton potential 
\cite{Copeland}.

\section{Constraints from CMB anisotropies}

The amplitude of scalar perturbations is given by \cite{Maartens}

\beq
\label{eq:AS}
A_{s}^2 \simeq \left . \left({512\pi\over 75 M_P^6}\right) {V^3
\over V^{\prime2}}\left[ 1 + {V \over 2\lambda} \right]^3
\right|_{k=aH}~~,
\eeq
where the right-hand side should be evaluated as the comoving scale 
equals the Hubble radius during inflation, $k=a H$. Thus the amplitude 
of scalar perturbations is increased
relative to the standard result at a fixed value of $\phi$ for a given
potential. 

In order to obtain the value of $\phi$ when scales corresponding to 
large-angle CMB anisotropies, as observed by COBE, left the 
Hubble radius during inflation, we take $N_\star\approx 55$ and 
$\phi_i=\phi_\star$ in Eq.~(\ref{eq:N}). Combining with 
Eq.~(\ref{eq:AS}) and using the fact that the  observed value 
from COBE is $A_{s}=2\times 10^{-5}$, we obtain, after a numerical analysis, 
a further (stronger) constraint on $\mu$

\beq
\mu \lsim 2 \times 10^{-8}~ M~~.
\label{eq:mu2}
\eeq
For $\mu=2 \times 10^{-8}~ M$, we get $\phi_\star=0.09~ 
M_P$ and $M_5=3\times 10^{-4}~ M_P$ 
(for $\phi_f=0.03~ M_P$, as obtained from Eq.~(\ref{eq:phif})).

The scale-dependence of the perturbations is described by the
spectral tilt \cite{Maartens}

\beq
n_{s}-1\equiv {d\ln A_{s}^2 \over d\ln k} \simeq -6\epsilon + 2\eta~~,
\label{eq:ns}
\eeq
where the slow-roll parameters are given in Eqs.~(\ref{eq:epsilon})
and~(\ref{eq:eta}).  Notice that, as
$V/\lambda\to\infty$, the spectral index is driven towards the
Harrison-Zel'dovich spectrum, $n_{s}\to 1$.
For our model, Eq.~(\ref{eq:pot}), we obtain

\beq
n_s\simeq 0.95 ~~\hbox{for}~~ \mu=2\times 10^{-8} M~~.
\eeq

The ratio between the amplitude of tensor and scalar
perturbations is given by \cite{Langlois}
\beq
{A_{t}^2 \over A_{s}^2} \simeq {3 M_P^2\over 16 \pi} 
\left({V^\prime\over V} \right)^2 {2 \lambda\over V}~~.
\label{eq:r}
\eeq
We get, for our model 
\beq
r \simeq 4 \pi{ A_T^2\over A_s^2} \simeq 0.22~~\hbox{for}
~~\mu=2\times 10^{-8}~~,
\eeq
which is consistent with current upper limits, $r < 0.4$ (see eg. \cite{DASI}).

For consistency, it should be verified that enough inflation can occur.
Indeed, for $\mu=2\times 10^{-8}$, we get $N=65$ for 
$\phi_i=0.094~M_P$ and a huge amount of e-foldings can be obtained 
for larger values of $\phi_i$ e.g. $N = 9.1 \times 10^6 $ for $\phi_i=0.5~M_P$.
\begin{table}[t]
\caption[par]{\label{tab:par} Relevant physical quantities, 
for different values of  $\mu$, for the simple $N=1$ Supergravity model of 
Eq.~(\ref{eq:pot}).}
\begin{ruledtabular}
\begin{tabular}{|c|c|c|c|c|c|}
$\mu/M $ & $\phi_\star/M_P$ & $M_5/M_P$ & $n_s$ & $r$ & $T_{RH} (\gev)$ \\
\hline 
$ 2\times 10^{-8}$ & $0.09$ & $3\times 10^{-4}$ & $0.949$  & $0.21$ & $7.7\times 10^5$ \\ 
\hline
$ 10^{-8}$ & $0.035$ & $1.2\times 10^{-4}$ & $0.953$  & $0.18$ & $2.7\times 10^5$ \\ 
\hline
$5\times 10^{-9}$ & $0.018 $ & $6\times 10^{-5}$ & $0.955$  & $0.17$ & $9.6\times 10^4$ \\ 
\hline
$ 10^{-9}$ & $0.0035$ & $1.2 \times 10^{-5}$ & $0.955$  & $0.17$ & $8.6\times 10^3$ \\  
\hline
$ 10^{-10}$ & $0.00035$ & $1.2\times 10^{-6}$ & $0.955$  & $0.17$ & $272$ \\  
\end{tabular}
\end{ruledtabular}
\end{table}

For smaller values of $\mu$ (see Table~\ref{tab:par}), we get
smaller values of $\phi_\star$ and $M_5$, but $n_s$ and $r$ change very little.
The number of e-folds of inflation increases for the same initial value 
of $\phi$ and $\phi_f$, calculated according to Eq.~(\ref{eq:phif}); 
for instance, for $\mu= 10^{-10}~M_P$, $\phi_i=0.094~M_P$ and $\phi_f=
9\times 10^{-5}~M_P$, we get $N=3 \times 10^{11}$, about five  orders of
 magnitude greater than 
the value obtained for the case $\mu = 2 \times 10^{-8}~M$ (see above).
Notice that the values of $n_s$ and $r$ exhibited in
 Table~\ref{tab:par}  are 
consistent with latest CMB data from DASI \cite{DASI}, BOOMERANG \cite{Boom} 
and MAXIMA \cite{Maxima}. 
However, for $\mu \lsim 1 \times 10^{-10}~M$, the reheating temperature 
becomes smaller than the typical value for the temperature 
of the electroweak phase transition, $T_{EW} \simeq 300~\gev$.
Thus, $\mu \lsim 1 \times 10^{-10}~M$, 
implies a premature breaking of the electroweak phase 
transition, from which we can extract a new bound on the 5-dimensional mass
scale (see Table~\ref{tab:par})
\beq
M_5 \gsim  1.3 \times 10^{-6}~M_P~~.
\label{eq:m5}
\eeq                                                                        
                              
Naturally, as field values become smaller, the  approximation which
consists of  expanding the  potential in $\phi/M$ and take just the first 
(quadratic) term becomes better; in this case, it is easy to obtain analytic
expressions for the relevant quantities, as it is reduces to the case 
analysed in Ref.~\cite{Maartens}.

\section{Conclusions}

Hence, we see that it is possible to sucessfully implemment 
chaotic inflation in a simple $N=1$ Supergravity model within the 
braneworld scenario, without need to fine tune the parameters of the 
potential. In fact, a single mass parameter, $\mu$, is needed, as in $N=1$ 
Supergravity new inflationary models \cite{Ross,Holman}.

An upper bound on $\mu$, $\mu \lsim~3.7 \times 10^{-6}~M$, is obtained 
from the requirement that sufficiently few gravitinos are regenerated in the
post-inflationary reheating epoch. However, a more severe upper bound is 
obtained from requiring adequate density fluctuations (both slope and 
amplitude) as observed by COBE, namely $\mu \lsim 2 \times 10^{-8}~M$ 
implying, for the 5-dimensional mass scale that 
$M_5 \lsim  3 \times 10^{-4}~M_P$. 
Requiring that the reheating temperature be greater than the 
typical temperature of the electroweak phase transition, 
$T_{EW} \simeq 300~\gev$, leads to upper bounds on these quantities, namely  
 $\mu \gsim~1.06 \times 10^{-10}~M$ and   
$M_5 \gsim 1.3 \times 10^{-6}~M_P$.   

Finally, we have shown that, remarkably, successful chaotic inflation 
can be achieved with sub-Planckian field values, thus avoiding the need 
to invoke hypothetic symmetries that would prevent the presence of higher 
order non-renormalisable terms in the superpotential. This last feature is 
important to prevent an overproduction of tensor perturbations in the CMB, a 
problem that can be particularly acute in chaotic inflation scenarios 
\cite{Enqvist}, 
and that,  ultimately due to the quadratic energy density term that 
appears in the Friedmann equation in the context of the braneworld scenario, 
is absent in our model.

\begin{acknowledgments}

\noindent
The authors acknowledge the partial support of FCT (Portugal)
under the grant POCTI/1999/FIS/36285.
\end{acknowledgments}


\end{document}